\documentstyle[prl,aps,multicol,psfig]{revtex}

\begin{document}
\draft
\tighten

\title{ Classical Frustration and Quantum Disorder in Spin-Orbital Models }

\author { Jan Zaanen }
\address{ Lorentz Institute for Theoretical Physics, Leiden University,
          P.O.B. 9506, NL-2300 RA Leiden, The Netherlands }
\author { Louis Felix Feiner }
\address{ Institute for Theoretical Physics, Utrecht University,
          Princetonplein 5, NL-3584 CC Utrecht, The Netherlands, and \\
          Philips Research Laboratories, Prof. Holstlaan 4,
          NL-5656 AA Eindhoven, The Netherlands }
\author { Andrzej M. Ole\'{s} }
\address{ Institute of Physics, Jagellonian University, Reymonta 4, 
          PL-30059 Krak\'ow, Poland, and \\
          Max-Planck-Institut f\"ur Festk\"orperforschung, Heisenbergstrasse 1,
          D-70569 Stuttgart, Federal Republic of Germany }
\date{\today}
\maketitle

\begin{abstract} 
Recently much attention is paid to the role of the orbital degrees of
freedom in transition metal oxides as it remains unclear whether they can
remain in a quantum disordered state at zero temperature. Discrete symmetry
of the orbital sector counteracts the quantum melting, but especially in 
doped systems there are signs of dynamical frustration involving the spin-, 
charge-, and orbital sector simultaneously. We discovered that even the 
simple Kugel-Khomskii model, describing $e_g$ degenerate Mott-insulators, 
is characterized by a point of perfect dynamical frustration on the
classical level, reached in the absence of Hund's rule and electron-phonon
couplings. This frustration is lifted on the quantum level, and the true
nature of the ground state is still unknown. At present there are
two proposals: the KCuF$_3$ phase, stabilized by an order-out-of-disorder
mechanism, or spin-orbital valence bond phases. It will be argued that
at least in the Cu based systems of this kind, the electron-phonon 
coupling is primarely responsible for 
driving the systems away from the special point in the phase diagram.
\end{abstract}

\pacs{71.27.+a, 74.72.-h, 75.10.-b, 64.60.-i}

\begin{multicols}{2}
\narrowtext

\section{Introduction}

The paradigm of correlated electron physics is based on the idea that
for a certain category of systems one better starts out with the electronic 
structure of the atoms, treating the delocalization of the electrons
in the solid as a perturbation. Any student of physics has to struggle 
through the theory of atomic multiplets, which is rather complicated 
because of the intricacies associated with orbital angular momentum.
At first sight it is therefore remarkable that these orbital degrees
of freedom are completely neglected in the main stream of correlated
electron physics. Recently the interest in `orbitals' has been reviving,
especially since they appear to be relevant in one way or another in
the colossal magnetoresistance (CMR) manganites. In the wake of this
development, questions are asked on the relevancy of these orbitals
in the context of seemingly settled problems like the metal-insulator
transition in V$_2$O$_3$ \cite{Bao98}. In this contribution we will review 
yet another recent development. Even in the Mott-insulating limit, where 
the physics simplifies considerably, the interplay of orbital and spin 
degrees of freedom poses a problem of principle.   

There are two limits where the role of orbital degeneracy is well 
understood: (i) The `band structure limit', which is 
based on the assertion that electron correlations can be neglected.
In any modern  local density approximation (LDA) band 
structure calculation, orbitals are fully taken into account on the one 
particle level, in so far as the atomic limit is of any relevance.
These translate into various bands, giving rise to multi-sheeted
fermi surfaces, etcetera. (ii) The localized, orbital and spin ordered case 
which we will refer to as the `classical limit'. In Mott-insulators, orbital
degrees of freedom acquire a separate existence in much the
same way as the spins of the electrons do. The orbitals can be 
parametrized by pseudospins and these form together with the physical spins a
low energy sector which is described by generalizations of the Heisenberg 
spin-Hamiltonian \cite{Jan93}. These are the spin-orbital models, like the 
Kugel-Khomskii (KK) model for $e_g$ degenerate cubic cuprates \cite{Kug82}. 
The `classically' ordered states, becoming exact in the limit of infinite 
dimensions ($d\rightarrow\infty$) and/or large (pseudo) spin ($S\rightarrow 
\infty$), define what is usually meant with orbital and spin order.

The question arises if there are yet other possibilities. We started to study 
this problem quite some time ago \cite{Crete}, well before the subject 
revived due to the manganites. Our motivation was actually related to a 
theoretical development flourishing in the 1980's: large $N$ theories 
\cite{Aue94}. By enlarging the symmetry, say from $SU(2)$ to $SU(N)$ with 
$N$ large, new saddle points (ordered states) appear which correspond to the
fluctuation dominated (non-perturbative) limit of the large $S$/large $d$ 
theories. For a single correlated impurity, orbital degeneracy leads in a 
natural way to these large $N$ notions. We asked the question if these large 
$N$ notions could become of relevance in lattice problems. We focussed on 
the simple problem of the $e_g$ Jahn-Teller degenerate Mott-insulator, 
rediscovering the KK Hamiltonian \cite{Kug82}. We tried to tackle this 
problem using the techniques invented by Arovas and Auerbach for the $SU(N)$ 
symmetric Heisenberg model \cite{Aro88}. We found that the $SU(4)$ symmetry 
is so badly broken that the large $N$ techniques were of little help, which
is another way of saying that the physics of the KK model is not 
controlled by large global symmetry. However, we did find a special 
approximate solution which revealed that the quantum fluctuations are
actually enhanced, and this motivated us to study these fluctuations in more 
detail starting from the large $S$ limit. In this process we discovered that
the enhancement of the fluctuations is due to the control exerted by
a point in parameter space which can be either called an infinite order
quantum-critical point, or a point of perfect {\em dynamical frustration} 
in the classical limit \cite{Fei97}.

This phenomenon will be discussed in the next section. It poses a
rather interesting theoretical problem. So much is clear that the
ground state degeneracy of the classical limit is lifted by
quantum fluctuations and the question is on the character of the
true ground state. As will be discussed, either the classical spin-orbital
order might survive, stabilized by an order-out-of-disorder mechanism,
or quantum-incompressible valence-bond like states might emerge. In
Section III the role of electron-phonon coupling will be addressed, 
emphasizing the rather counter-intuitive result of LDA+U electronic
structure calculations that phonons play a rather secondary role
despite the fact that the lattice deformations are large. 
Finally, the situation in the manganites will be shortly
discussed in Section IV.

\section{ The Kugel-Khomskii model and dynamical frustration }

Consider a Mott-insulator which is characterized by orbital degeneracy, 
besides the usual spin degeneracy. Different from pure spin problems, these 
spin-orbital problems are rather ungeneric and depend on the precise system 
under consideration. A simple problem is a cubic lattice of $3d$-ions in a
$d^9$ configuration: the Kugel-Khomskii problem, which directly
applies to Cu perovskites like KCuF$_3$ or K$_2$CuF$_4$ \cite{Kug82}. The
large Mott gap in the charge excitation spectrum simplifies matters
considerably and one derives an effective Hamiltonian by insisting
on one hole per unit cell, deriving superexchange-like couplings
between the spin and orbital degrees of freedom by integrating
out virtual charge fluctuations.

The spins are described as usually in terms of an $su(2)$ algebra
($\vec{S}_i$). The orbital degrees of freedom are the $e_g$ cubic harmonics 
$x^2-y^2 \sim |x\rangle$ and $3z^2-r^2 \sim |z\rangle$, which can be
parametrized in terms of pseudospins as   
$|x\rangle ={\scriptsize\left( \begin{array}{c} 1\\ 0\end{array}\right)},\; 
 |z\rangle ={\scriptsize\left( \begin{array}{c} 0\\ 1\end{array}\right)}$. 
Pauli matrices $\sigma^u$ ($u = x, y, z$) are introduced acting on these
states. Different from the spins, the $SU(2)$ symmetry associated with the 
pseudospins is badly broken because the orbitals communicate with the 
underlying lattice. Although the $e_g$ states are degenerate on a single 
site, this degeneracy is broken by the virtual charge fluctuations, which
take place along the interatomic bonds, i.e., in a definite direction
with respect to the orientation of the orbitals. It is therefore convenient 
to introduce operators which correspond to orbitals directed either along 
or perpendicular to the three cubic axes $\alpha=a,b,c$, given by 
$(\tau^{\alpha}_j-\frac{1}{2})$ and $(\tau^{\alpha}_j+\frac{1}{2})$, where 
\begin{equation} 
\tau^{a(b)}_i =\frac{1}{4}( -\sigma^z_i\pm\sqrt{3}\sigma^x_i ), 
\hskip 1cm
\tau^c_i = \frac{1}{2} \sigma^z_i \;.
\label{orbop}
\end{equation}
In terms of these operators, the Kugel-Khomskii Hamiltonian can be written 
as ($J=t^2/U$ and $t$ is the hopping along the $c$-axis) \cite{Fei97}, 
\begin{eqnarray} 
\label{kk1}
H_1 = &J& \sum_{\langle ij\rangle,\alpha} \left[ 4(\vec{S}_i\cdot\vec{S}_j ) 
  (\tau^{\alpha}_i - \frac{1}{2}) (\tau^{\alpha}_j - \frac{1}{2})\right.
                                   \nonumber \\ 
& & \hskip 1.0cm + \left. (\tau^{\alpha}_i+\frac{1}{2})(\tau^{\alpha}_j 
+ \frac{1}{2}) - 1 \right] ,
\end{eqnarray} 
neglecting the Hund's rule splittings $\propto J_H$ of the intermediate 
$d^8$ states ($J_H$ is the singlet-triplet splitting).
Including those up to order $\eta=J_H/U$ yields in addition,
\begin{eqnarray} 
\label{kk2}
H_2 = & J\eta & \sum_{\langle ij\rangle,\alpha}
  \left[ (\vec{S}_i\cdot\vec{S}_j) 
  (\tau^{\alpha}_i + \tau^{\alpha}_j - 1 )  \right.   \nonumber \\
&+& \left. \frac{1}{2}(\tau^{\alpha}_i-\frac{1}{2}) 
                      (\tau^{\alpha}_j-\frac{1}{2}) 
 + \frac{3}{2} (\tau^{\alpha}_i \tau^{\alpha}_j - \frac{1}{4})\right] . 
\end{eqnarray}
Eq.'s (\ref{kk1},\ref{kk2}) are rather unfamiliar: they describe a regular
Heisenberg spin problem coupled into a Pott's like orbital problem (choose
two out of three possibilities $\sim x^2-y^2, \sim y^2-z^2, \sim z^2-x^2$).

The oddity of Eq.'s (\ref{kk1},\ref{kk2}) becomes clear when one studies
the classical limit. As usually, the $\vec{S}$'s and the $\vec{\tau}$'s
are treated as classical vectors. In order to draw a phase diagram 
we introduced another control parameter,
\begin{equation} 
\label{kk3}
H_3 = - E_z \sum_i \tau^z_i,
\end{equation}
a "magnetic field" for the orbital pseudo-spins, loosely associated with 
a uniaxial pressure along the $c$-axis. The classical limit phase diagram 
as function of $\eta$ and $E_z$ is shown in Fig. 1.

\begin{figure} \unitlength1cm
\begin{picture}(8,8)
\put(0.,0.){\psfig{figure=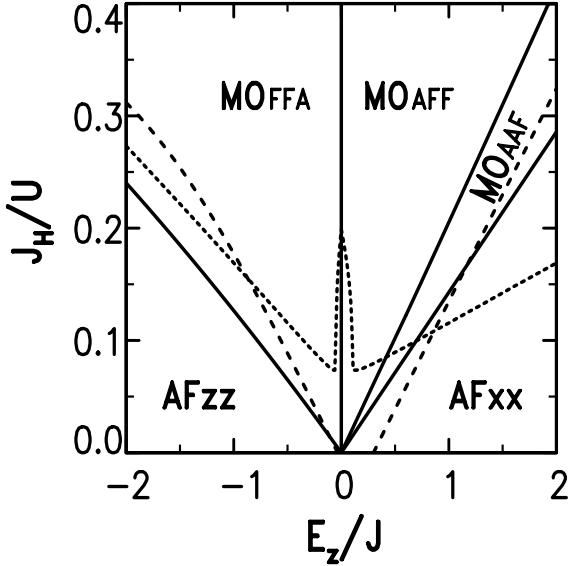,height=7.5cm,width=7.5cm,angle=0}}
\end{picture}
\caption{ Phase diagram of the Kugel-Khomskii model in the classical limit,
as function of the Hund's rule coupling $J_H$ and tetragonal crystal field 
$E_z$ (reproduced from Ref. \protect\cite{Fei97}).}   
\label{f1}
\end{figure}

For a detailed discussion of the various phases we refer to Ref. \cite{Fei97}.
To give some feeling, for large positive $E_z$ the $x^2-y^2$ orbitals are 
occupied, forming $(a,b)$ planes of antiferromagnetically coupled spins 
(AFxx). This is nothing else than the situation realized in, e.g. 
La$_2$CuO$_4$. For large negative $E_z$ the $3z^2-r^2$ orbitals condense, 
forming a 3D spatially anisotropic Heisenberg antiferromagnet [AFzz with 
stronger exchange coupling along the $c$-axis than in the $(a,b)$ planes]. 
Finally, the MOFFA, MOAFF and MOAAF phases are variations of the basic 
Kugel-Khomskii spin-orbital order \cite{Kug82} obtained by rotating the 
magnetic and orbital structure by $\pi/2$. For the MOFFA phase at $E_z =0$, 
the orbitals have a two-sublattice structure in the $(a,b)$-planes ($x^2-z^2$ 
and $y^2-z^2$ on the A- and B-sublattice, respectively). Along the $c$-axis 
strong antiferromagnetic spin-spin couplings are found, while the spin 
couplings in the $(a,b)$ planes are ferromagnetic with a strength $\sim\eta$.   
 
The anomaly occurs at the origin $(E_z,\eta)=(0,0)$ of the phase diagram:
a 3D antiferromagnet (AFzz), a 2D antiferromagnet (AFxx) and a quasi-1D
A-type antiferromagnet (MOFFA/MOAFF/MOAAF) become degenerate! The emphasis on
the `uniaxial pressure' $E_z$ is misleading in the sense that the full
scope of the problem is not visible directly from this phase diagram: at
the origin of Fig. 1 an {\em infinity\/} of classical phases become
degenerate. This is trivial to understand. In the absence of Hund's rule
exchange, the Hamiltonian Eq. (\ref{kk1}) becomes the full story. 
Assuming a 3D classical antiferromagnet, $\vec{S}_i \cdot \vec{S}_j = -1/4$,
and inserting this in Eq. (\ref{kk1}) yields,
\begin{equation}
H_{eff} = J\sum_{\langle ij\rangle,\alpha} 
\left( \tau_i^{\alpha} + \tau_j^{\alpha} - 1 \right) .
\label{3ddeg}
\end{equation}
The orbital degrees of freedom are completely decoupled and all $2^N$ orbital 
configurations have the same energy ($\sum_{\alpha}\tau_i^{\alpha}=0$)!
In addition, this infinity of different 3D spin systems has the same energy 
as the MOFFA/MOAFF/MOAAF phases. It is actually so that at any finite 
temperature the 3D antiferromagnet becomes stable because of the entropy 
associated with the decoupled orbital sector \cite{janup}.

This `gauge' degeneracy is clearly a pathology of the classical limit. We
continued by studying the stability of the classical phase diagram with
respect to Gaussian quantum fluctuations. As discussed in more detail in Ref. 
\cite{Foz98} this is a somewhat subtle affair. Intuitively, one could be 
tempted to think that the orbitals and spins can be excited independently.
This is however not the case. The dynamical algebra of relevance to the 
problem is an $so(4)$ algebra, and this implies that modes will occur which
excite at the same time the spins and the orbitals: the spin-and-orbital 
waves (SOW)'s.
Next to a (longitudinal) sector of pure orbital excitations, a `transversal'
sector is found corresponding with spin-excitations which are mixed with
spin-and-orbital excitations, except for the acoustic modes at long
wavelength which become pure spin-waves as imposed by the Goldstone theorem.  

We found that upon approaching the infinite critical point, the mass gap
associated with the discrete symmetry in the orbital sector collapses.      
The (mixed) transverse modes give the dominating contribution to the 
renormalization of energy and magnetic order parameter. In the AFxx (AFzz) 
phase the lowest transverse mode softens along $\vec{k}=(\pi,0,k_z)$ 
[$\vec{k}=(k_x,0,0)$], and equivalent lines in the Brillouin zone (BZ), 
regardless how one approaches the critical lines. Thus, these modes become 
dispersionless along particular (soft-mode) lines in the BZ, where we find 
{\em finite\/} masses in the perpendicular directions,
\begin{eqnarray} 
\omega_{\rm AFxx}(\vec{k}) \rightarrow & \Delta_x &
      + B_x \left( k_x^4 + 14k_x^2k_y^2 + k_y^4 \right)^{1/2}, \nonumber \\
\omega_{\rm AFzz}(\vec{k}) \rightarrow & \Delta_z & 
      + B_z \left( k_y^2 + 4k_z^2 \right), 
\label{mass0}
\end{eqnarray}
with $\Delta_i=0$ and $B_i\neq 0$ at the $M$ point, and the quantum 
fluctuations diverge logarithmically, $\langle\delta S^z\rangle\sim
\int d^3k/\omega(\vec{k})\sim\int d^2k/(\Delta_i+B_ik^2)\sim\ln\Delta_i$, 
if $\Delta_i\rightarrow 0$ at the transition. We found that the quantum
correction to the order parameter $\langle S^z\rangle$ becomes large,
well before the critical point is reached. In Fig. 1 the lines are
indicated where  $|\langle \delta S^z\rangle|=\langle S^z\rangle$:
in the area enclosed by the dashed and dotted lines classical order
cannot exist, at least not in gaussian order. 
 
If the classical limit is as sick as explained in the previous 
paragraphs, what is happening instead? {\it A priori\/} it is not 
easy to give an answer to this question. There are no `off the shelf'
methods to treat quantum spin problems characterized by classical 
frustration, and the situation is similar to what is found in, e.g. 
$J_1-J_2-J_3$ problems \cite{Pre88}. A first possibility is quantum 
order-out-of-disorder \cite{Chu91}: quantum fluctuations can stabilize 
a particular classical state over other classically degenerate states, if 
this particular state is characterized by softer excitations than any of the 
other candidates. Khaliullin and Oudovenko \cite{Kha97} have suggested that
this mechanism is operative in the present context, where the AFzz
3D anisotropic antiferromagnet is the one becoming stable. Their original
argument was flawed because of the decoupling procedure they used, which
violates the $so(4)$ dynamical algebra constraints \cite{Foz98}. However, 
Khaliullin claims to have found an `$so(4)$ preserving' self-consistent 
decoupling procedure which does yield order-out-of-disorder \cite{Kha98}. 
Nevertheless, there is yet another possibility: valence-bond (VB) singlet 
(or spin-Peierls) order, which at the least appears in a more natural way 
in the present context than is the case in higher dimensional spin-only 
problems, because it is favored by the directional nature of the orbitals.

The essence of a (resonating) valence bond [(R)VB] state is that one 
combines pairs of spins into singlets. In the short-range (R)VB states these 
singlets involve nearest-neighbor spin pairs. Subsequently, one particular
covering of the lattice with these `spin-dimers' might be favored 
(VB or spin-Peierls state), or the ground state might become a coherent 
superposition of many of these coverings (RVB state). On a cubic lattice the 
difficulty is that although much energy is gained in the formation of the 
singlet pairs, the bonds between the singlets are treated poorly. 
Nevertheless, both in 1D spin systems (Majumdar-Ghosh \cite{Maj69}, 
AKLT-systems \cite{Aff87}) and in the large $N$ limit of $SU(N)$ magnets in 
2D, ground states are found characterized by spin-Peierls/VB order \cite{Read}. 

It is straightforward to understand that the interplay of orbital- and spin 
degrees of freedom tends to stabilize VB order. Since the orbital sector is 
governed by a discrete symmetry, the orbitals
tend to condense in some classical orbital order. Different from
the fully classical phases, one now looks for orbital configurations
optimizing the energy of the spin VB configurations. The spin energy
is optimized by having orbitals $3\zeta^2-r^2$ on the nearest-neighbor
sites where the VB spin-pair lives, with $\zeta$ directed along the bond. 
This choice maximizes the overlap between the wave functions, and thereby the 
binding energy of the singlet. At the same time, this choice of orbitals 
minimizes the unfavorable overlaps with spin pairs located in directions 
orthogonal to $\zeta$. The net result is that VB states are much better 
variational solutions for the KK model, as compared to the standard 
Heisenberg spin systems.

\begin{figure} \unitlength1cm
\begin{picture}(8,8)
\put(0.,0.){\psfig{figure=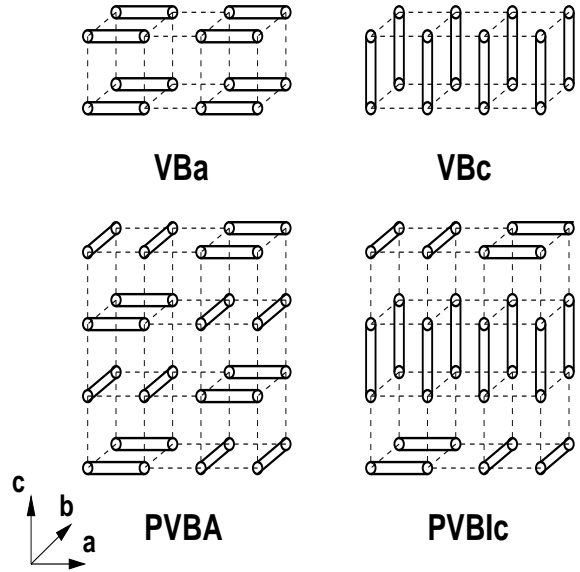,height=7.5cm,width=7.5cm,angle=0}}
\end{picture}
\caption{ A variety of valence bond solids (see text). }  
\label{f2}
\end{figure}

Adressing this systematically, we found that two families of VB states
are most stable: (i) The `staggered' VB states like the PVBA and 
PVBIc states of Fig. 2. These states have in common that the overlap
between neighboring VB pairs is minimized: the large lobes of the
$3\zeta^2-r^2$ wave functions of different pairs are never pointing
to each other. (ii) The `columnar' VB states like the VBc (or VBa) 
state of Fig. 2. In the orbital sector, this is nothing else than
the AFzz state of Fig. 1 ($3z^2-r^2$ orbitals on every site). Different
from the AFzz state, the spin system living on this orbital backbone is 
condensed in a 1D spin-Peierls state along the $z$-direction which is
characterized by strong exchange couplings. The spins in the 
$a(b)$-directions stay uncorrelated, due to the weakness of the respective 
exchange couplings as compared to the VB mass gap.

The energies of these VB states and the classical states dressed up with 
quantum fluctuations are quite close together. A key issue is if the true 
ground state is compressible (dressed classical state), or characterized 
by a dynamical mass-gap (VB states). This will most likely depend on
subtleties beyond the reach of the relatively crude variational Ans\"atze 
presented here \cite{notekhalu}. So the nature of the ground state of the 
Kugel-Khomskii problem for small Hund's-rule coupling is still an open 
problem.

\section{ Electron-phonon coupling in KC\lowercase{u}F$_3$ }

In the previous Section we discussed the orbital order as driven by
the electron-electron interactions. However, one can think quite
differently about the real systems: the deformations found in 
KCuF$_3$ (or LaMnO$_3$) could in principle be entirely caused by
phonon-driven collective Jahn-Teller effects. This subject has 
been intensely studied in the past and is well understood. 
It starts out neglecting electron-electron interactions,
and the focus is instead on the electron-phonon coupling. In case
that the ions are characterized by a Jahn-Teller (orbital) degeneracy,
one can integrate out the (optical) phonons, and one finds effective
Hamiltonians with phonon mediated interactions between the orbitals.
In the specific case of $e_g$ degenerate ions in a cubic crystal, these 
look quite similar to the KK Hamiltonian, except that the spin dependent
term is absent\cite{KKphon}. Any orbital order resulting from this 
Hamiltonian is now accompanied by a lattice distortion of the same symmetry.

The size of the quadrupolar deformation in the $(a,b)$
plane of KCuF$_3$ is actually as large as 4 \% of the lattice constant ($a$).
It is therefore often argued that the orbital order is clearly phonon-driven, 
and that the physics of the previous section is an irrelevancy. Although
appealing at first sight, this argument is flawed: large displacements
do not necessarily imply that phonons do all the work. 

The deformations of the lattice and the orbital degrees of freedom cannot 
be disentangled using general principles: they constitute an irreducible 
subsector of the problem. The issue is therefore a quantitative one, and 
in the absence of experimental guidance one would therefore like to address 
the issue with a quantitative electronic structure method. The LDA+U method 
is the method of choice. It is constructed to handle the physics of 
electronic orbital ordering, keeping the accurate treatment of the 
electron-lattice interaction of LDA intact. According to LDA+U calculations 
the total energy gained by the deformation of the lattice is minute as 
compared to the energies involved in the electronic orbital ordering
\cite{Lie95}. At the same time, the phonons are important on the macroscopic 
scale and they contribute to driving KCuF$_3$ away from the infinite-critical 
point of the phase diagram Fig. 1.

We start out with the observation
that according to LDA KCuF$_3$ would be an undistorted, cubic
system: the energy increases if the distortion is switched on (see
Fig. 3). The reason is that KCuF$_3$ would be a band metal according to
LDA (the usual Mott-gap problem) with a Fermi-surface which is not
susceptible to a band Jahn-Teller instability. LDA+U yields a drastically
different picture \cite{Lie95}. LDA can be looked at as unpolarized LDA+U, 
and by letting both the orbitals and the spins polarize an energy is gained 
of order of the band gap, i.e., of the order of 1 eV. The orbital- and
spin polarization is nearly complete and the situation is close to the 
strong coupling limit underlying the spin-orbital models of Section II.
Also when the cubic lattice is kept fixed, the correct orbital and spin
ordering (MOFFA of Fig. 1) is found, with spin-exchange constants which 
compare favorably with experiment \cite{Lie95}. Because the orbital
order has caused the electron density to become highly unsymmetric, 
the cubic lattice is unstable. Further energy can be gained by letting the 
lattice relax. The lattice distortion calculated in LDA+U ($\sim$ 3\% of $a$) 
comes close to the actual distortion of KCuF$_3$ ($\sim$ 4 \%).
However, despite the fact that the distortion is large, the energy gained by 
the lattice relaxation is rather minute: $\sim 50$ meV (see Fig. 3)! 
Obviously, in the presence of the electronic orbital order the cubic lattice 
becomes very soft with regard to the quadrupolar distortions and even a small 
electron-phonon coupling can cause large distortions.

\begin{figure} \unitlength1cm
\begin{picture}(8,8)
\put(0.,0.){\psfig{figure=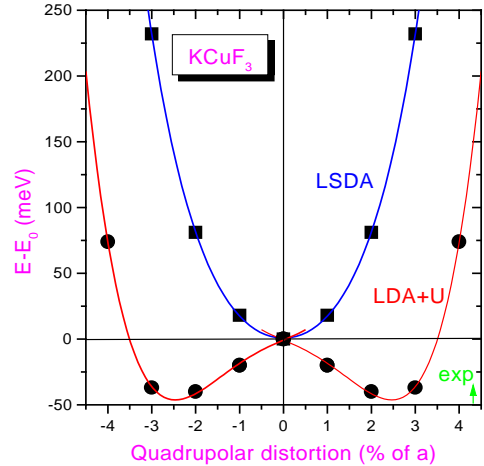,height=7.5cm,width=7.5cm,angle=0}}
\end{picture}
\caption{The dependence of the total energy of KCuF$_3$ on the
quadrupolar lattice distortion according to LSDA and LDA+U band
structure calculations (after Ref. \protect\cite{Lie95}).}
\label{f3}
\end{figure}

Although the energy gained in the deformation of the lattice is rather
small, the electron-phonon coupling is quite effective in keeping KCuF$_3$ 
away from the physics associated with the origin of the phase diagram (Fig. 
1). Since the ferromagnetic interactions in the $(a,b)$ plane of KCuF$_3$ 
are quite small ($J_{ab}=-0.2$ meV, as compared to the `1D' exchange 
$J_c=17.5$ meV \cite{Ten95}), one might argue that the effective Hund's rule 
coupling $J\eta$ as of relevance to the low energy theory is quite small.  
Although this still needs further study, it might well be that in the absence 
of the electron-phonon coupling KCuF$_3$ would be close to the origin of Fig. 
1. However, the electron-phonon coupling can be looked at as yet another axis 
emerging from the origin. In principle, the electron-phonon coupling 
introduces two scales: (i) a retardation scale, which is governed by the 
ratio of the phonon frequency and the electronic scale set by $J\sim 20$ meV. 
Since $J$ is relatively small, KCuF$_3$ is close to the anti-adiabatic limit
where the lattice follows the electronic fluctuations, (ii) in
the anti-adiabatic limit the phonons are high energy modes which can be
integrated out, causing the effective orbital-orbital couplings we earlier
referred to. These couplings destroy the cancellations leading to Eq. 
(\ref{3ddeg}), thereby driving the system away from the point of classical 
degeneracy. The typical scale for the phonon induced effective orbital 
interactions is at most of the order of the LDA+U lattice relaxation energy. 
However, as the latter ($\sim 50$ meV) is quite a bit larger than $J$, the 
effective interaction will likely be able to put KCuF$_3$ well outside the 
`dangerous' region near the origin of the phase diagram.

In summary, although further work is needed it might be that phonons are
to a large extent responsible for the stability of KCuF$_3$'s classical 
ground state. In any case, one cannot rely on the sheer size
of the lattice deformations to resolve this issue!

\section{How about the manganites ?}

Given the discussion so far, the search for interesting quantum effects
in orbital degenerate Mott-insulators should not be regarded as hopeless. 
Unfortunately, the insulating parent compounds of the CMR manganites, such 
as LaMnO$_3$, are {\it not\/} candidates for this kind of physics. The 
reason is not necessarily phonons: also in the manganites the `Jahn-Teller' 
lattice distortions are sizable, but this does not necessarily imply that 
the phonons are dominating. Two of us derived a Kugel-Khomskii-type model 
of relevance to this regime, and we did find a dynamical frustration of 
$e_g$-superexchange at $J_H\simeq 0$ \cite{Fei98}. However, the system is 
driven away from this point by two effects:
 (i) the manganites are in the Hund's rule dominated regime, with a large 
     splitting between the lowest energy high-spin state at $U-5J_H$ 
     (with $J_H=0.69$ eV \cite{Miz95}), and 
     the low-spin states at energies $\sim U$;
(ii) the additional $t_{2g}$-superexchange between the $S=3/2$ cores favours 
     an antiferromagnetic order in all three spatial directions.
The net outcome is that the ferromagnetic
interaction between the {\em total $S=2$ spins\/} in the $(a,b)$ planes 
is of order of the $c$-axis exchange, signalling that the manganites are in
the Hund's rule stabilized regime of the phase diagram. 

The mysteries of the manganites relate to what happens when 
quantum-mechanical holes are added to the orbital/spin ordered insulator. 
This is undoubtedly a problem with its own characteristics, which cannot 
be reduced to a variation on the far simpler problems encountered in the 
insulators. Nevertheless, we do believe that the study of the insulating 
limit might be of some help in better appreciating what is going on in the 
doped systems.
It is tempting to think about orbital degrees of freedom as 
being spins in disguise. This is not quite the case. Orbitals are far less 
quantum-mechanical -- they are more like Ising spins than Heisenberg spins. 
Secondly, orbitals carry this unfamiliar property that depending on their 
specific orientation in internal space, overlaps increase in particular real 
space directions, while they diminish in orthogonal directions. 
Our valence-bond constructions illustrate this peculiar phenomenon in the 
case of spins, but the same logic is at work when the hole is delocalizing. 
This intimate connection between internal symmetry and the directionality of 
delocalization causes the dynamical frustration which has been highlighted 
in this communication. This motive seems also at work in the doped system, 
witness the many near degenerate states found both in mean-field 
calculations \cite{Miz95,Nag98} and in experiment \cite{Tok94}. 
Further work is needed on this fascinating problem.

{\it Acknowledgements}. We thank A. I. Lichtenstein for helpful discussions.
AMO acknowledges support by the Committee of Scientific Research (KBN) of
Poland, Project No. 2 P03B 175 14.

\end{multicols}


\begin{references}
\bibitem{notekhalu} For instance, one can argue that our columnar VB states 
                    are quite like the `order-out-of-disorder' states of Ref. 
                    \cite{Kha97}, with the only difference that we have
                    imposed a spin-Peierls order in the chain direction.
\end{references}
\end{document}